\documentclass[a4paper]{article}

\usepackage{indentfirst}
\usepackage{amsmath}
\usepackage[english]{babel}
\usepackage{graphicx}
\usepackage{subfigure}
\usepackage{multirow}
\usepackage{longtable}
\usepackage{fullpage}
\usepackage{float}
\usepackage[section]{placeins}
\usepackage{authblk}

\title{Landau theory of nuclear level density and its application in description of nuclear level density in the region of discrete and s-wave neutron resonance energies}

\author[]{A. Elmas}
\author[]{H. Ahmadov\thanks{Corresponding author: ahmadov@gantep.edu.tr}}
\author[]{B. Gonul}
\affil[]{Department of Engineering Physics, University of Gaziantep, 27310 Gaziantep, T\"{u}rkiye}

\date{\today}

\begin{document}

\maketitle

\begin{abstract}

In this work, the reliability of the Landau expression for the nuclear level density calculations is tested, for the first-time, to describe nuclear level densities of some light, intermediate mass and heavy nuclei at excitations corresponding to discrete and s-wave neutron resonance energies. The $\chi^{2}$ minimizing method is used in treatment of the experimental data for the two suggested energy range of discrete energies given by Nuclear Data Sheet \cite{ripl3} and by the systematic for nuclear level density parametrization in \cite{koning2008}. Our comparison with the related data in the discrete energy range has shown that the results obtained by the Landau expression  are better than those of back-shifted Fermi-gas model and constant temperature approximation. This result is also valid for some nuclei of interest when the s-wave neutron resonance level density is included to check theoretical prescriptions in the energy range from initial bound states to unbound states near the neutron binding energy. 
\vspace{5mm}

\noindent
\textit{Keyword(s):} Level density, s-wave neutron resonance, Landau theory

\end{abstract}

\section{Introduction}
\label{sec:1}

Nuclear level density (LD) has been the subject of many theoretical and experimental investigations such as analyses of fusion and fission reactions for a wide range of isotopes and reactor structure designs. Investigations of the problems related to nuclear LD  also clarifies some basic properties of the nuclear matter such as super-fluidity, nuclear shell, pairing and collective effects.

Up to date, the independent particle model of Fermi gas (FG) model after the Bethe's LD expression \cite{bethe1936} and its various modifications \cite{gc1965,ignatyuk1979,dilg1973} are frequently used in describing the corresponding experimental data including the LD of discrete energy levels and s-wave neutron resonance densities, as well as in the description of nuclear reaction product particle spectra [7-13]. The so-called composite Gilbert-Cameron (GC) model  \cite{gc1965,koning2008,voinov2013} unifies further FG model with the constant temperature (CT) approximation. These models however do not involve the collective excitation mechanism of the nucleus in the form of collective vibrations and rotation effects in an explicit manner, despite these effects may be described somehow by the LD parameter ($a$) and the concerning energy shift ($\Delta$). The explicit collective effect contributions to LD's are well described, for instance, in the works of Refs. \cite{ignatyuk1979,ignatyuk1993} and in the LD systematic given in \cite{koning2008}. Apart from such effects, in these works the LD parameter ($a$) is supposed to be excitation energy ($U$) dependent. 

The need for the inclusion of an energy dependent LD parameter arises due to shell effects in the LD descriptions. When this term is used in the generalized super-fluid (GS) nuclear model , the related parameter $a$ changes the shape of LD function above the critical energy $U_c$, characterizing transitions from super-fluid to normal state. While the microscopic models of  nuclear LD includes both Hartee-Fock-Bogolyubov (HFB) model of nuclear structure and combinatorial model analysis of LD's \cite{hilaire2006}.

All the models mentioned above aim to describe the observed LD's accurately in a wide range of isotopes and excitation energies, which is not yet accomplished. Therefore introduction of new models and theories in describing LD's is still required in theoretical nuclear physics.

Within this context, the motivation behind the present work is to test the earlier LD expression of Landau theory of a nucleus  described by \cite{landau1937} (later abbreviated by L) for the theoretical analysis of nuclear LD's in the region of both discrete nuclear  energy range and at the nuclear excitations corresponding to the s-wave neutron resonance energies. To our knowledge, such an investigation has never been appeared so far in the related literature.

Section \ref{sec:2} gives a necessary brief theoretical background of Landau's theory of nuclear LD and description of the models used through the present work. Comparison of the model findings is the subject of Sections \ref{sec:3} and \ref{sec:4}. Finally, some concluding remarks are given in Section \ref{sec:5}. 

\section{Nuclear level density of Landau's theory}
\label{sec:2}
In this theory nuclear system is considered as "quantum liquid" of nucleon system in which interaction between the particles is significant.  
According to the statistical theory at low excitation energies, nuclear free energy ($F$) dependence on nuclear temperature ($T$) may be expressed as \cite{landau1937}

\begin{equation}
\label{eq:1}
\operatorname{F}(T)=\frac{1}{2}\frac{\partial^2 F}{\partial T^2}\bigg|_{T=0} T^2\quad.
\end{equation}
In the above equation, normally nuclear ground states are admitted to be with zero excitation and the derivation uses Nernst theorem, that is the property of quantum systems, and accordingly linear dependence on $T$ is dropped. Introducing  $\alpha\equiv-\frac{\partial^2 F}{\partial T^2}\big|_{T=0}$ entropy ($S$) and excitation energy ($U$) may be written as
 
\begin{equation}
\label{eq:2}
S=-\frac{\partial F}{\partial T}=\alpha T,\quad U=F+TS=\frac{\alpha T^2}{2}\quad.
\end{equation}
The number of nuclear states ($N$) and the nuclear state density ($\rho$) of L theory are expressed as
\begin{equation}
\label{eq:3}
N=e^{\operatorname{S}(U)}=e^{\sqrt{2 \alpha U}},\quad \operatorname{\rho}(U)=\frac{dN}{dU}=\frac{1}{T}e^{\sqrt{2 \alpha U}}\quad.
\end{equation} 
The spin ($J$) dependent nuclear LD formula then with a great accuracy appears as (\cite{landau1937})

\begin{equation}
\label{eq:4}
\operatorname{\rho}(U,J)=\frac{(2J+1)e^{-\frac{(J+\frac{1}{2})^2}{2 \sigma^2}}}{2 \sqrt{2 \pi} \sigma^3}\frac{e^{\sqrt{2 \alpha U}}}{T}\quad,
\end{equation}
where

\begin{equation}
\label{eq:5}
\sigma^2=\frac{I_0}{\hbar^2}T
\end{equation}
with $I_0$ ($=\frac{2}{5}MR^2$) being the rigid body moment of inertia of spherical nuclei in which $R=r_0A^{\frac{1}{3}}$ and $M=m_nA$ together with $m_n$ being the nucleon mass while $A$   denotes the mass of nuclei. The parameter $\sigma$ denotes the spin cutoff parameter, more specifically, the rigid body (RB) spin cutoff parameter. Further, introducing a new parameter $a\equiv\frac{\alpha}{2}$, which is identified as LD parameter of L theory, we have the traditional well-known expression for the excitation energy

\begin{equation}
\label{eq:6}
U=aT^2\quad,
\end{equation}
and substituting Eq.\eqref{eq:6} into Eq.\eqref{eq:5} and calculating $\frac{I_0}{\hbar^2}$ for $ r_0=1.5fm$ one finds,

\begin{equation}
\label{eq:7}
\sigma^2=0.0214A^{5/3}\bigg(\frac{U}{a}\bigg)^{1/2}\quad,
\end{equation}
where $U$ and $a$ are expressed in terms of MeV's. Eq.\eqref{eq:4} may be easily generalized for different models as

\begin{equation}
\label{eq:8}
\operatorname{\rho}(U,J)=\frac{(2J+1)e^{-(J+1/2)^2/{2\sigma^2}}}{2\sqrt{2\pi}\sigma^3}\operatorname{\rho_{th}}(U)
\end{equation}
in which $\rho_{th}$ is the theoretical or model state density function of interest. Note that this expression is applicable for each parity. 
The total LD defined as in the work of \cite{gc1965} 

\begin{equation}
\label{eq:9}
\operatorname{\rho}(U)=\sum_{\substack{J,\Pi}}(2J+1)\operatorname{\rho}(U,J,\Pi)\quad,
\end{equation}
includes however, excitation energy ($U$), spin ($J$) and parity ($\Pi$) dependent level density, $\operatorname{\rho}(U,J,\Pi)$. For equal distribution of parities $\operatorname{\rho}(U,J,\Pi)$ is expressed as

\begin{equation}
\label{eq:10}
\operatorname{\rho}(U,J,\Pi)=\frac{1}{2}\operatorname{\rho}(U,J)\quad.
\end{equation}
Substituting Eq.\eqref{eq:10} and Eq.\eqref{eq:8}in Eq.\eqref{eq:9}, summing over $\Pi$, and replacing the summation over $J$ in Eq.\eqref{eq:9} by the related integration one find that

\begin{equation}
\label{eq:11}
\operatorname{\rho}(U)=\operatorname{\rho_{th}}(U)\quad,
\end{equation}
while, $(1/{2\sqrt{2\pi}\sigma^3})\int_{0}^{\infty}(2J+1)e^{-(J+1/2)^2/{2\sigma^2}}dJ=1$. Hence, CT approximation, FG model and L in Eq.\eqref{eq:11} can be expressed as

\begin{equation}
\label{eq:12}
\rho_{CT}=\frac{1}{T}\exp\bigg(\frac{U}{T}\bigg)\quad,
\end{equation}
\begin{equation}
\label{eq:13}
\rho_{FG}=\frac{\sqrt{\pi}}{12}\frac{\exp(2\sqrt{aU})}{a^{1/4}U^{5/4}}\quad,
\end{equation}
\begin{equation}
\label{eq:14}
\rho_L=\frac{\exp(2\sqrt{a_L U})}{T_L}\quad.
\end{equation}
$T$ in Eq.\eqref{eq:12} is the constant temperature parameter for a given nucleus whereas in Eq.\eqref{eq:14} $T_L$ is given by Eq.\eqref{eq:6} and, similarly, $a$ and $a_L$ are the LD parameters of FG and L, respectively. We shall use these Eqs.\eqref{eq:12}-\eqref{eq:14} in our analysis of the related experimental LD's in Section 4. In the fitting procedure to the experimental data in the domain of discrete and resonance energies, the shift ($\Delta$) to the excitation energy $U$, $U\rightarrow U-\Delta $ is employed and the parameters $a$ and $\Delta$ are considered as adjustable parameters within the frame of Eq.\eqref{eq:8}, while the spin cutoff parameters in the L and FG treatments are assumed as behaving either in the form of Eq.\eqref{eq:7} or in the form of the so-called GC spin cutoff \cite{gc1965} given by

\begin{equation}
\label{eq:15}
\sigma^2=0.0888A^{2/3}(aU)^{1/2}
\end{equation}
unlike the CT approximation where the spin cutoff parameter is used as in Ref.\cite{egidy2005}

\begin{equation}
\label{eq:16}
\sigma=0.98A^{0.29}\quad.
\end{equation}

In the next section(Section 3) we study mass dependence of resonance LD parameter and excitation energy behavior of the LD of L theory in comparison with the same properties of other models. 

\section{Comparison of the resonance level density parameters}
\label{sec:3}

To predict the resonance LD parameters of different prescriptions used here, the observed s-wave neutron resonance LD(data, taken from \cite{ripl3}), $\frac{1}{D_{res}}$ is fitted to the resonance LD model prescriptions $\operatorname{\rho_{res}}$, including equal parity distribution of the levels for the nucleus considered with the spins $J=I-1/2$ and $J=I+1/2$ (where $I$ is the target spin) at the neutron binding energy. Model resonance level density is expressed as

\begin{eqnarray}
\label{eq:17}
\operatorname{\rho_{res}}&=&\frac{1}{2}[\operatorname{\rho}(U,I-1/2)+\operatorname{\rho}(U,I+1/2)]\nonumber\\
&\cong& \frac{(2I+1)}{2\sqrt{2\pi}\sigma^3}e^{-(I+1/2)^2/{2\sigma^2}}\operatorname{\rho_{th}}(U)
\end{eqnarray}
Note that, $D_{res}$ is the average s-resonance spacing while $\sigma$, for calculation of this section is taken as in Eq.\eqref{eq:7}. This expression form enables us to make a clear and reliable comparison considering our predictions on $a_{res}$ of L with those of FG, together with the additional model function

\begin{equation}
\label{eq:18}
\operatorname{\rho}(U)=\frac{\exp(2\sqrt{aU})}{\sqrt{48}U}
\end{equation}
which is the version of FG model for the identical fermions(IF), protons or neutrons \cite{bethe1936}. Eq. \eqref{eq:18} has a different excitation energy dependence for the denominator comparing with denominators of Eqs.\eqref{eq:13} and \eqref{eq:14}, which serves as a benchmark test of the present Landau theory. $T_{res}$ parameters for CT model can be found by substituting Eq.\eqref{eq:12} into the Eq.\eqref{eq:17}.  Through this section, the nuclear excitation energy U is taken as $U\rightarrow U-\delta$ where $\delta$ is the sum of neutron and proton pairing energies taken from \cite{gc1965,baba1970}.

\begin{table}[!htbp] %disc tablosu
\centering
\scalebox{0.75}
{
\begin{tabular}{|c|c|c|c|c|c|c|c|c|}
\hline
\multirow{2}{*}{\parbox{2.2cm}{\centering Compound \\ Nuclei}} & \multicolumn{1}{c|}{I} & \multicolumn{1}{c|}{$B_{n}$} & \multicolumn{1}{c|}{$\delta$} & \multicolumn{1}{c|}{$D_{obs}$} & \multicolumn{1}{c|}{$a_{res}(FG)$} & \multicolumn{1}{c|}{$a_{res}(IP)$} & \multicolumn{1}{c|}{$a_{res}(L)$} & \multicolumn{1}{c|}{$T_{res}(CT)$}\\
&  & \multirow{1}{*}{(MeV)} & \multirow{1}{*}{(MeV)} & \multirow{1}{*}{(keV)} & \multirow{1}{*}{$(MeV^{-1})$} & \multirow{1}{*}{$(MeV^{-1})$} & \multirow{1}{*}{$(MeV^{-1})$} & \multirow{1}{*}{(MeV)} \\
\hline
\multirow{1}{*}{\parbox{2.4cm}{\centering\textsuperscript{24}Na}} & 3/2 & 6.96 & 0 & 100 & 3.33 & 2.77 & 1.09 & 1.21 \\
\hline
\multirow{1}{*}{\parbox{2.4cm}{\centering \textsuperscript{25}Mg}} & 0 & 7.331 & 2.46 & 480 & 3.55 & 2.87 & 0.80 & 0.96 \\
\hline
\multirow{1}{*}{\parbox{2.4cm}{\centering \textsuperscript{32}P}} & 1/2 & 7.937 & 0 & 50 & 4.02 & 3.36 & 1.37 & 1.12 \\
\hline
\multirow{1}{*}{\parbox{2.4cm}{\centering \textsuperscript{36}Cl}} & 3/2 & 8.58 & 0 & 23 & 4.15 & 3.50 & 1.56 & 1.15 \\
\hline
\multirow{1}{*}{\parbox{2.4cm}{\centering \textsuperscript{42}K}} & 3/2 & 7.534 & 0 & 10 & 5.51 & 4.68 & 1.19 & 0.92 \\
\hline
\multirow{1}{*}{\parbox{2.4cm}{\centering \textsuperscript{44}Ca}} & 7/2 & 11.131 & 3.27 & 1.8 & 7.13 & 6.19 & 3.37 & 0.80 \\
\hline
\multirow{1}{*}{\parbox{2.4cm}{\centering \textsuperscript{49}Ti}} & 0 & 8.142 & 1.73 & 20.8 & 6.92 & 5.88 & 2.71 & 0.75 \\
\hline
\multirow{1}{*}{\parbox{2.4cm}{\centering \textsuperscript{52}V}} & 7/2 & 7.311 & 0 & 4 & 6.80 & 5.87 & 3.09 & 0.81 \\
\hline
\multirow{1}{*}{\parbox{2.4cm}{\centering \textsuperscript{55}Cr}} & 0 & 6.246 & 1.35 & 50 & 7.48 & 6.29 & 2.68 & 0.64 \\
\hline
\multirow{1}{*}{\parbox{2.4cm}{\centering \textsuperscript{58}Fe}} & 1/2 & 10.044 & 2.83 & 7.05 & 6.95 & 5.94 & 2.88 & 0.79 \\
\hline
\multirow{1}{*}{\parbox{2.4cm}{\centering \textsuperscript{62}Ni}} & 3/2 & 10.596 & 2.61 & 2.1 & 7.15 & 6.16 & 3.17 & 0.81 \\
\hline
\multirow{1}{*}{\parbox{2.4cm}{\centering \textsuperscript{64}Cu}} & 3/2 & 7.916 & 0 & 0.7 & 8.47 & 7.35 & 3.96 & 0.73 \\
\hline
\multirow{1}{*}{\parbox{2.4cm}{\centering \textsuperscript{67}Zn}} & 0 & 7.052 & 1.06 & 4.62 & 9.83 & 8.47 & 4.33 & 0.59 \\
\hline
\multirow{1}{*}{\parbox{2.4cm}{\centering \textsuperscript{70}Ga}} & 3/2 & 7.654 & 0 & 0.316 & 9.80 & 8.54 & 4.75 & 0.66 \\
\hline
\multirow{1}{*}{\parbox{2.4cm}{\centering \textsuperscript{74}Ge}} & 9/2 & 10.196 & 3.24 & 0.062 & 12.78 & 11.28 & 6.78 & 0.53 \\
\hline
\multirow{1}{*}{\parbox{2.4cm}{\centering \textsuperscript{76}As}} & 3/2 & 7.328 & 0 & 0.09 & 12.00 & 10.53 & 6.10 & 0.57 \\
\hline
\multirow{1}{*}{\parbox{2.4cm}{\centering \textsuperscript{81}Se}} & 0 & 6.701 & 1.43 & 2.4 & 12.40 & 10.73 & 5.67 & 0.49 \\
\hline
\multirow{1}{*}{\parbox{2.4cm}{\centering \textsuperscript{86}Rb}} & 5/2 & 8.650 & 0 & 0.172 & 9.44 & 8.27 & 4.74 & 0.70 \\
\hline
\multirow{1}{*}{\parbox{2.4cm}{\centering \textsuperscript{88}Sr}} & 9/2 & 11.112 & 2.17 & 0.29 & 8.62 & 7.57 & 4.40 & 0.75 \\
\hline
\multirow{1}{*}{\parbox{2.4cm}{\centering \textsuperscript{91}Zr}} & 0 & 7.195 & 1.20 & 6 & 9.89 & 8.52 & 4.36 & 0.59 \\
\hline
\multirow{1}{*}{\parbox{2.4cm}{\centering \textsuperscript{95}Zr}} & 0 & 6.462 & 1.20 & 4 & 11.81 & 10.19 & 5.30 & 0.51 \\
\hline
\multirow{1}{*}{\parbox{2.4cm}{\centering \textsuperscript{96}Mo}} & 5/2 & 9.154 & 2.40 & 0.081 & 12.98 & 11.40 & 6.64 & 0.53 \\
\hline
\multirow{1}{*}{\parbox{2.4cm}{\centering \textsuperscript{100}Tc}} & 9/2 & 6.764 & 0 & 0.014 & 15.66 & 13.89 & 8.59 & 0.46 \\
\hline
\multirow{1}{*}{\parbox{2.4cm}{\centering \textsuperscript{102}Ru}} & 5/2 & 9.219 & 2.22 & 0.018 & 14.98 & 13.25 & 8.04 & 0.49 \\
\hline
\multirow{1}{*}{\parbox{2.4cm}{\centering \textsuperscript{106}Pd}} & 5/2 & 9.561 & 2.59 & 0.0103 & 15.99 & 14.18 & 8.72 & 0.47 \\
\hline
\multirow{1}{*}{\parbox{2.4cm}{\centering \textsuperscript{108}Ag}} & 1/2 & 7.271 & 0 & 0.028 & 15.28 & 13.53 & 8.25 & 0.49 \\
\hline
\multirow{1}{*}{\parbox{2.4cm}{\centering \textsuperscript{114}Cd}} & 1/2 & 9.043 & 2.68 & 0.0248 & 17.54 & 15.53 & 9.48 & 0.43 \\
\hline
\multirow{1}{*}{\parbox{2.4cm}{\centering \textsuperscript{115}Sn}} & 0 & 7.546 & 1.19 & 0.286 & 14.46 & 12.70 & 7.39 & 0.48 \\
\hline
\multirow{1}{*}{\parbox{2.4cm}{\centering \textsuperscript{119}Sn}} & 0 & 6.484 & 1.19 & 0.7 & 15.31 & 13.37 & 7.52 & 0.44 \\
\hline
\multirow{1}{*}{\parbox{2.4cm}{\centering \textsuperscript{123}Te}} & 0 & 6.929 & 1.14 & 0.146 & 17.09 & 15.06 & 8.92 & 0.42 \\
\hline
\multirow{1}{*}{\parbox{2.4cm}{\centering \textsuperscript{125}Sn}} & 0 & 5.733 & 1.19 & 5 & 13.49 & 11.63 & 6.02 & 0.44 \\
\hline
\multirow{1}{*}{\parbox{2.4cm}{\centering \textsuperscript{126}Te}} & 1/2 & 9.113 & 2.23 & 0.043 & 15.59 & 13.78 & 8.34 & 0.48 \\
\hline
\multirow{1}{*}{\parbox{2.4cm}{\centering \textsuperscript{130}I}} & 7/2 & 6.463 & 0 & 0.02 & 16.02 & 14.18 & 8.62 & 0.45 \\
\hline
\multirow{1}{*}{\parbox{2.4cm}{\centering \textsuperscript{134}Cs}} & 7/2 & 6.891 & 0 & 0.021 & 15.09 & 13.35 & 8.13 & 0.48 \\
\hline
\multirow{1}{*}{\parbox{2.4cm}{\centering \textsuperscript{137}Ba}} & 0 & 6.906 & 1.58 & 1.21 & 14.46 & 12.60 & 6.99 & 0.45 \\
\hline
\multirow{1}{*}{\parbox{2.4cm}{\centering \textsuperscript{139}La}} & 5 & 8.778 & 0.85 & 0.032 & 12.73 & 11.28 & 6.90 & 0.56 \\
\hline
\multirow{1}{*}{\parbox{2.4cm}{\centering \textsuperscript{142}Pr}} & 5/2 & 5.843 & 0 & 0.11 & 14.80 & 12.98 & 7.51 & 0.46 \\
\hline
\multirow{1}{*}{\parbox{2.4cm}{\centering \textsuperscript{148}Pm}} & 7/2 & 5.895 & 0 & 0.0052 & 20.20 & 17.97 & 11.24 & 0.38 \\
\hline
\multirow{1}{*}{\parbox{2.4cm}{\centering \textsuperscript{150}Sm}} & 7/2 & 7.987 & 2.21 & 0.0024 & 22.21 & 19.81 & 12.58 & 0.36 \\
\hline
\multirow{1}{*}{\parbox{2.4cm}{\centering \textsuperscript{153}Sm}} & 0 & 5.868 & 1.22 & 0.046 & 24.10 & 21.35 & 13.03 & 0.31 \\
\hline
\multirow{1}{*}{\parbox{2.4cm}{\centering \textsuperscript{156}Gd}} & 3/2 & 8.536 & 1.89 & 0.0017 & 21.22 & 18.97 & 12.20 & 0.39 \\
\hline
\multirow{1}{*}{\parbox{2.4cm}{\centering \textsuperscript{163}Dy}} & 0 & 6.271 & 0.92 & 0.062 & 20.73 & 18.35 & 11.18 & 0.36 \\
\hline
\multirow{1}{*}{\parbox{2.4cm}{\centering \textsuperscript{167}Er}} & 0 & 6.436 & 0.62 & 0.038 & 20.29 & 18.00 & 11.12 & 0.38 \\
\hline
\multirow{1}{*}{\parbox{2.4cm}{\centering \textsuperscript{169}Yb}} & 0 & 6.867 & 0.68 & 0.008 & 22.34 & 19.95 & 12.75 & 0.37 \\
\hline
\multirow{1}{*}{\parbox{2.4cm}{\centering \textsuperscript{174}Yb}} & 5/2 & 7.464 & 1.23 & 0.0075 & 19.09 & 16.96 & 10.56 & 0.41 \\
\hline
\multirow{1}{*}{\parbox{2.4cm}{\centering \textsuperscript{176}La}} & 7/2 & 6.289 & 0 & 0.00345 & 20.11 & 17.93 & 11.35 & 0.39 \\
\hline
\multirow{1}{*}{\parbox{2.4cm}{\centering \textsuperscript{177}Hf}} & 0 & 6.383 & 0.64 & 0.03 & 21.13 & 18.78 & 11.67 & 0.37 \\
\hline
\multirow{1}{*}{\parbox{2.4cm}{\centering \textsuperscript{180}Hf}} & 9/2 & 7.388 & 1.37 & 0.0046 & 20.23 & 18.02 & 11.37 & 0.38 \\
\hline
\multirow{1}{*}{\parbox{2.4cm}{\centering \textsuperscript{182}Ta}} & 7/2 & 6.063 & 0 & 0.0042 & 20.45 & 18.22 & 11.48 & 0.38 \\
\hline
\multirow{1}{*}{\parbox{2.4cm}{\centering \textsuperscript{187}W}} & 0 & 5.467 & 0.72 & 0.099 & 22.28 & 19.68 & 11.85 & 0.33 \\
\hline
\multirow{1}{*}{\parbox{2.4cm}{\centering \textsuperscript{190}Os}} & 3/2 & 7.792 & 1.88 & 0.0034 & 22.51 & 20.08 & 12.76 & 0.36 \\
\hline
\multirow{1}{*}{\parbox{2.4cm}{\centering \textsuperscript{196}Pt}} & 1/2 & 7.922 & 1.50 & 0.0018 & 23.55 & 21.11 & 13.75 & 0.36 \\
\hline
\multirow{1}{*}{\parbox{2.4cm}{\centering \textsuperscript{202}Hg}} & 3/2 & 7.754 & 1.58 & 0.09 & 15.52 & 13.67 & 8.06 & 0.46 \\
\hline
\multirow{1}{*}{\parbox{2.4cm}{\centering \textsuperscript{204}Tl}} & 1/2 & 6.656 & 0 & 0.47 & 12.92 & 11.32 & 6.47 & 0.53 \\
\hline
\multirow{1}{*}{\parbox{2.4cm}{\centering \textsuperscript{206}Tl}} & 1/2 & 6.503 & 0 & 4 & 9.98 & 8.63 & 4.56 & 0.62 \\
\hline
\multirow{1}{*}{\parbox{2.4cm}{\centering \textsuperscript{208}Pb}} & 1/2 & 7.368 & 1.21 & 30 & 7.66 & 6.53 & 3.09 & 0.71 \\
\hline
\multirow{1}{*}{\parbox{2.4cm}{\centering \textsuperscript{209}Pb}} & 0 & 3.937 & 0.83 & 90 & 12.45 & 10.49 & 4.57 & 0.40 \\
\hline
\multirow{1}{*}{\parbox{2.4cm}{\centering \textsuperscript{210}Bi}} & 9/2 & 4.604 & 0 & 4.45 & 11.23 & 9.67 & 4.99 & 0.50 \\
\hline
\multirow{1}{*}{\parbox{2.4cm}{\centering \textsuperscript{230}Th}} & 5/2 & 6.794 & 1.38 & 0.00062 & 28.07 & 25.17 & 16.45 & 0.30 \\
\hline
\multirow{1}{*}{\parbox{2.4cm}{\centering \textsuperscript{233}Th}} & 0 & 4.786 & 0.78 & 0.0165 & 31.86 & 28.37 & 17.83 & 0.25 \\
\hline
\multirow{1}{*}{\parbox{2.4cm}{\centering \textsuperscript{234}U}} & 5/2 & 6.845 & 1.26 & 0.00052 & 27.74 & 24.89 & 16.33 & 0.31 \\
\hline
\multirow{1}{*}{\parbox{2.4cm}{\centering \textsuperscript{235}U}} & 0 & 5.298 & 0.69 & 0.0112 & 29.16 & 26.01 & 16.53 & 0.28 \\
\hline
\multirow{1}{*}{\parbox{2.4cm}{\centering \textsuperscript{237}U}} & 0 & 5.126 & 0.69 & 0.014 & 29.57 & 26.35 & 16.66 & 0.27 \\
\hline
\multirow{1}{*}{\parbox{2.4cm}{\centering \textsuperscript{238}Np}} & 5/2 & 5.488 & 0 & 0.00057 & 28.01 & 25.12 & 16.45 & 0.31 \\
\hline
\multirow{1}{*}{\parbox{2.4cm}{\centering \textsuperscript{241}Pu}} & 0 & 5.242 & 0.61 & 0.013 & 28.68 & 25.57 & 16.21 & 0.28 \\
\hline
\multirow{1}{*}{\parbox{2.4cm}{\centering \textsuperscript{244}Am}} & 5/2 & 5.367 & 0 & 0.00073 & 28.02 & 25.12 & 16.38 & 0.30 \\
\hline
\end{tabular}
}

\caption{Comparison of calculated resonance LD parameters of different nuclei.}
\label{table:1}
\end{table}

\begin{figure}[ht] %ht here top demek
\centering
\subfigure[]{\includegraphics[width=0.41\linewidth]{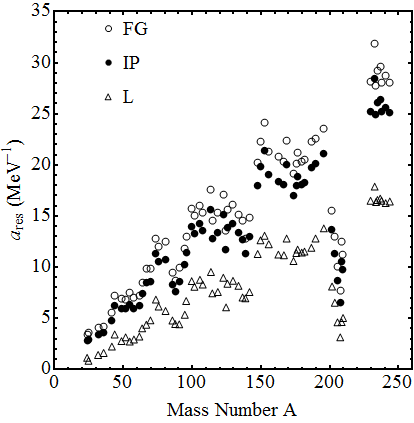}}
\label{fig:1a}
\quad
\subfigure[]{\includegraphics[width=0.43\linewidth]{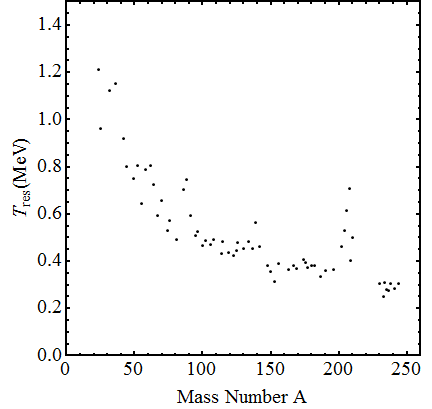}}
\label{fig:1b}
\caption{(a) Comparison of the resonance LD parameters $a_{res}$ for different compound nuclei obtained using LD's expressed by Eqs.\eqref{eq:13},\eqref{eq:14} and \eqref{eq:18} for the RB spin cutoff. (b) $T_{res}$ parameter values of CT model for different compound nuclei obtained using LD expressed by Eq.\eqref{eq:12}}.
\label{fig:1}
\end{figure}
 The findings of the calculations underlined for different nuclei are illustrated in Table 1 and in Figs. 1a and 1b. As can readily be seen from Figs. 1a and 1b considering all the model calculations, shell effects are explicitly appearing near the doubly magic nuclei. The resonance LD parameter values for FG and IP models are very close to each other, whereas the LD parameters obtained by the present L theory are systematically lower than those of two mentioned models. This effect is due to the energy dependencies of denominators of corresponding equations for the LD's. If this dependence express as $\propto{U^k}$, then ${k=1/2}$ for L, and ${k=1}$ and ${k=5/4}$ for IP and FG, respectively. Slower energy dependence in denominator result smaller value of $a_{res}$.  Clearly this feature must also appear in the fitting of the models to the corresponding experimental data including both the discrete energies and s-wave neutron excitations. The finding $T_{res}$ parameter values of CT model decrease versus the mass number of the compound nucleus however, indicating resonance increase near the double magic nuclei. 
\section{Fitting procedure for the models used and the results obtained}
\label{sec:4}

In the present work, experimental data for the two sets of discrete level numbers are used in the fitting of model functions for LD's. The first set includes the level numbers indicated by the RIPL-3 data sheet where energy levels up to maximum of the range are to be assumed as completely determined. This set of data are used in the fitting of discrete levels alone and also when discrete and resonance data are fitted with model LD's. The second set of the level numbers is taken from the systematic of the work in \cite{koning2008}, where the less numbers of levels comparing with RIPL-3 data are chosen. This data set for the discrete level numbers are used in the fitting procedure of discrete and resonance data with the model functions. The s-wave neutron resonance spacings are taken from RIPL-3 data.

\subsection{Discrete energy range}
\label{sec:4.1}

 We first approximate level number up to excitation energy $U$ in the discrete energy range by the use of formula

\begin{equation}
\label{eq:19}
\operatorname{N}(U)=\int_{0}^{U}\operatorname{\rho^{\prime}}(U)dU=\operatorname{N}(U_{min})+\int_{U_{min}}^{U}\operatorname{\rho^{\prime}}(U)dU\quad,
\end{equation}
where $\operatorname{N}(U_{min})$ is the number of level corresponding to the level energy $U_{min}$, $\rho^{\prime}=\rho_{CT}$ for the CT approximation (Eq. \eqref{eq:12}), and $\rho^{\prime}=\frac{1}{\sqrt{2\pi}\sigma}\rho_{th} $ for the FG and L (Eqs. \eqref{eq:13} and \eqref{eq:14}). Energy shift $\Delta$ to the excitation energy $U$ is employed in all formulas. The two versions of spin cutoffs given by the Eqs. \eqref{eq:7} (RB) and \eqref{eq:15} (GC) are used in the fitting procedures of both FG and L. For the CT the analytical solution of Eq.\eqref{eq:19}

\begin{equation}
\label{eq:20}
\operatorname{N}(U)=\operatorname{N}(U_{min})+(e^{\frac{U}{T}}-e^{\frac{U_{min}}{T}})e^{-\frac{\Delta}{T}}
\end{equation}
is used.

The fit of level numbers to the experimental RIPL-3 data for the  \textsuperscript{60}Co, \textsuperscript{62}Ni, \textsuperscript{76}As, \textsuperscript{116}Sn, \textsuperscript{138}Ba, \textsuperscript{139}La, \textsuperscript{166}Ho, \textsuperscript{198}Au, \textsuperscript{233}Th nuclei, which are the examples of deformed and spherical nuclei with even and odd nucleon numbers, are made by the use of the model parameters found by the minimization of

\begin{equation}
\label{eq:21}
\chi^2=\sum_{i=N_{min}}^{N_{max}}\bigg(\frac{\operatorname{N}(U_i)-i}{\sqrt{i}}\bigg)^2\quad,
\end{equation}
where $i$ is the level's number starting from $N_{min}$ and ending with $N_{max}$ while $U_i$ and $\sqrt{i}$ are the energy and error of $i$'th level number, respectively. Table 2 illustrates the parameter values obtained in the discrete level number fitting procedure of the CT, FG and L to the RIPL-3 data. $\chi^2$ values indicate that L theory in general describes the discrete level numbers versus excitation energy better than those of CT and FG models. Eq.\eqref{eq:20} is used further, in the description of discrete and unbound states. Figs. 2a and 2b show a comparison of the model fits of discrete states of the nuclei \textsuperscript{60}Co and \textsuperscript{198}Au when RIPL-3 data are used for their level schemes.

\begin{table}[!htbp] %disc tablosu
\centering
\scalebox{0.96}{
\begin{tabular}{|c|c|c|c|c|c|c|c|c|c|c|c|}
\hline
\multirow{2}{*}{\parbox{2.2cm}{\centering Nuclei\\Data Range}} &\multicolumn{3}{c|}{CT} & \multicolumn{4}{c|}{FG} & \multicolumn{4}{c|}{L}\\
& $T$ & $\Delta$ & $\chi^2$ & $\sigma$ & $a$ & $\Delta$ & $\chi^2$ & $\sigma$ & $a_L$ & $\Delta$ & $\chi^2$ \\
\hline
\multirow{2}{*}{\parbox{2.4cm}{\centering\textsuperscript{60}Co\\10-65}} & 1.04 & -1.97 & 1.78 & GC & 6.51 & -1.53 & 1.67 & GC & 3.26 & 0.04 & 1.57 \\
\cline{5-12}
&  &  &  & RB & 6.71 & -1.80 & 1.67 & RB & 3.60 & -0.41 & 1.58 \\
\hline
\multirow{2}{*}{\parbox{2.4cm}{\centering \textsuperscript{62}Ni\\10-49}} & 1.09 & 0.00 & 6.55 & GC & 4.00 & -2.13 & 5.49 & GC & 2.06 & 0.56 & 5.50 \\
\cline{5-12}
&  &  &  & RB & 4.20 & -2.87 & 5.49 & RB & 2.33 & -0.40 & 5.50 \\
\hline
\multirow{2}{*}{\parbox{2.4cm}{\centering \textsuperscript{76}As\\2-82}} & 1.33 & -5.44 & 23.65 & GC & 5.60 & -4.43 & 23.33 & GC & 3.03 & -2.33 & 23.05 \\
\cline{5-12}
&  &  &  & RB & 5.84 & -4.91 & 23.34 & RB & 3.36 & -2.97 & 23.11 \\
\hline
\multirow{2}{*}{\parbox{2.4cm}{\centering \textsuperscript{116}Sn\\10-45}} & 0.41 & 1.87 & 1.04 & GC & 15.96 & 1.88 & 1.09 & GC & 7.88 & 2.52 & 0.97 \\
\cline{5-12}
&  &  &  & RB & 13.82 & 1.48 & 1.62 & RB & 7.65 & 2.20 & 1.37 \\
\hline
\multirow{2}{*}{\parbox{2.4cm}{\centering \textsuperscript{138}Ba\\4-50}} & 1.21 & -1.21 & 3.18 & GC & 5.68 & -0.98 & 3.29 & GC & 2.87 & 0.84 & 3.39 \\
\cline{5-12}
&  &  &  & RB & 6.06 & -1.67 & 3.89 & RB & 3.35 & 0.01 & 3.37 \\
\hline
\multirow{2}{*}{\parbox{2.4cm}{\centering \textsuperscript{139}La\\4-33}} & 0.60 & 0.00 & 13.64 & GC & 3.59 & -5.54 & 2.73 & GC & 1.91 & -2.37 & 2.72 \\
\cline{5-12}
&  &  &  & RB & 3.88 & -6.93 & 2.73 & RB & 2.25 & -3.93 & 2.73 \\
\hline
\multirow{2}{*}{\parbox{2.4cm}{\centering \textsuperscript{166}Ho\\11-154}} & 0.84 & -3.42 & 15.54 & GC & 9.17 & -2.82 & 15.02 & GC & 5.10 & -1.45 & 14.62 \\
\cline{5-12}
&  &  &  & RB & 9.64 & -3.20 & 15.05 & RB & 5.70 & -1.91 & 14.72 \\
\hline
\multirow{2}{*}{\parbox{2.4cm}{\centering \textsuperscript{198}Au\\13-91}} & 1.38 & -5.89 & 1.45 & GC & 5.56 & -5.03 & 1.45 & GC & 3.08 & -2.79 & 1.45 \\
\cline{5-12}
&  &  &  & RB & 5.98 & -5.93 & 1.45 & RB & 3.57 & -3.80 & 1.45 \\
\hline
\multirow{2}{*}{\parbox{2.4cm}{\centering \textsuperscript{233}Th\\3-65}} & 1.02 & -3.97 & 4.02 & GC & 7.34 & -3.57 & 4.03 & GC & 3.97 & -1.96 & 4.04 \\
\cline{5-12}
&  &  &  & RB & 7.87 & -4.20 & 4.03 & RB & 4.61 & -2.68 & 4.03 \\
\hline
\end{tabular}
}
\caption{The parameters of the LD models obtained by $\chi^2$ minimizing of level numbers for RIPL-3. The related level numbers used are shown via the first column below the nucleus symbol.}
\label{table:2}
\end{table}
\begin{figure}[!htbp] %ht here top demek
\centering
\subfigure[\textsuperscript{60}Co]{\includegraphics[width=0.43\linewidth]{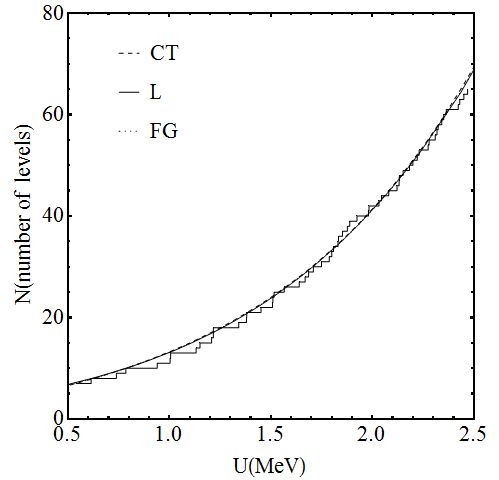}}
\label{fig:2a}
\quad
\subfigure[\textsuperscript{198}Au]{\includegraphics[width=0.43\linewidth]{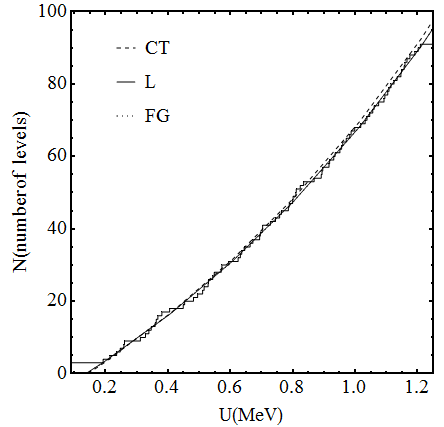}}
\label{fig:2b}
\caption{Staircase graphs of level numbers and their fit with different LD model functions a) \textsuperscript{60}Co, level numbers 10-65. b) \textsuperscript{198}Au, level numbers 13-91. The GC spin cutoff parameter is used for the FG and L.}
\label{fig:2}
\end{figure}

\subsection{Level numbers in the range of discrete and unbound states around the neutron binding energy}
\label{sec:4.2}

Parameter values of the model fits in the energy range is derived in minimizing of the relation

\begin{equation}
\label{eq:22}
\chi^2=\sum_{i=N_{min}}^{N_{max}}\bigg(\frac{\operatorname{N}(U_i)-i}{\sqrt{i}}\bigg)^2+\bigg(\frac{\rho_{obs}-\rho_{res}}{\Delta \rho_{obs}}\bigg)^2\quad,
\end{equation}
where $\rho_{obs}$ is the density of s-wave neutron resonance levels with the spin $I-1/2$ and $I+1/2$ determined by $\rho_{obs}=\frac{1}{D_{obs}}$ with  $D_{obs}$ being the average s-wave resonance spacing. In the above equation $\rho_{res}$ is the density of s-wave neutron resonance levels with spins $I-1/2$ and $I+1/2$ given by Eq. \eqref{eq:17}, $\Delta\rho_{obs}=\frac{1}{\Delta D_{obs}}$ where $\Delta D_{obs}$ is the error in average level spacing, and $\operatorname{N}(U_i)$ is calculated from Eq. \eqref{eq:19} for each of $\rho_{th}$ given by Eqs. \eqref{eq:12}, \eqref{eq:13}, \eqref{eq:14}.

\begin{table}[!htbp]
\centering
\scalebox{0.85}{
\begin{tabular}{|c|c|c|c|c|c|c|c|c|c|c|c|}
\hline
\multirow{2}{*}{Nuclei} & \multirow{2}{*}{Data Range} & \multirow{2}{*}{$\sigma$}  & \multicolumn{3}{|c|}{CT} & \multicolumn{3}{c|}{FG} & \multicolumn{3}{c|}{L}\\
&   &   &  $T$ & $\Delta$ &  $\chi^2$   &  $a$  &  $\Delta$  & $\chi^2$  & $a_L$   & $\Delta$  &  $\chi^2$ \\
\hline
\multirow{6}{*}{\textsuperscript{60}Co} & \multirow{3}{*}{10-65 (RIPL-3) + res.} & Eq.\eqref{eq:16} & 1.17 & -2.54 & 2.31 & - & - & - & - & - & - \\
\cline{3-12}
& & GC & - & - & - & 6.96 & -1.26 & 1.92 & 3.94 & 0.47 & 4.23 \\
\cline{3-12}
& & RB & - & - & - & 7.59 & -1.24 & 2.66 & 4.61 & 0.32 & 7.43 \\
\cline{2-12}
& \multirow{3}{*}{7-15 (Ref.\cite{koning2008}) + res.} & Eq.\eqref{eq:16} & 1.13 & -2.13 & 0.11 & - & - & - & - & - & - \\
\cline{3-12}
& & GC & - & - & - & 7.03 & -1.17 & 0.11 & 3.96 & 0.42 & 0.12 \\
\cline{3-12}
& & RB & - & - & - & 7.62 & -1.22 & 0.11 & 4.60 & 0.18 & 0.14 \\
\hline
\multirow{6}{*}{\textsuperscript{62}Ni} &  \multirow{3}{*}{10-49 (RIPL-3) + res.} & Eq.\eqref{eq:16} & 1.31 & -0.99 & 5.81 & - & - & - & - & - & - \\
\cline{3-12}
& & GC & - & - & - & 5.62 & -0.01 & 6.05 & 2.99 & 1.99 & 6.31 \\
\cline{3-12}
& & RB & - & - & - & 6.81 & 0.45 & 6.99 & 4.24 & 2.30 & 8.90 \\
\cline{2-12}
& \multirow{3}{*}{6-18 (Ref.\cite{koning2008}) + res.} & Eq.\eqref{eq:16} & 1.29 & -0.75 & 1.61 & - & - & - & - & - & - \\
\cline{3-12}
& & GC & - & - & - & 5.67 & 0.09 & 1.53 & 2.99 & 2.01 & 1.47 \\
\cline{3-12}
& & RB & - & - & - & 6.77 & 0.38 & 1.41 & 4.18 & 2.10 & 1.23 \\
\hline
\multirow{6}{*}{\textsuperscript{76}As} & \multirow{3}{*}{2-82 (RIPL-3) + res.} & Eq.\eqref{eq:16} & 0.95 & -3.43 & 31.31 & - & - & - & - & - & - \\
\cline{3-12}
& & GC & - & - & - & 9.73 & -1.92 & 58.40 & 5.43 & -0.66 & 79.69 \\
\cline{3-12}
& & RB & - & - & - & 10.70 & -1.81 & 74.46 & 6.33 & -0.71 & 119.39 \\
\cline{2-12}
& \multirow{3}{*}{6-27 (Ref.\cite{koning2008}) + res.} & Eq.\eqref{eq:16} & 0.95 & -3.51 & 1.74 & - & - & - & - & - & - \\
\cline{3-12}
& & GC & - & - & - & 9.90 & -1.98 & 1.48 & 5.80 & -0.67 & 1.34 \\
\cline{3-12}
& & RB & - & - & - & 10.85 & -1.89 & 1.41 & 7.04 & -0.64 & 1.23 \\
\hline
\multirow{6}{*}{\textsuperscript{116}Sn} &  \multirow{3}{*}{10-45 (RIPL-3) + res.} & Eq.\eqref{eq:16} & 0.72 & 0.65 & 4.50 & - & - & - & - & - & - \\
\cline{3-12}
& & GC & - & - & - & 13.54 & 1.62 & 1.59 & 7.91 & 2.52 & 1.05 \\
\cline{3-12}
& & RB & - & - & - & 14.95 & 1.65 & 1.25 & 9.75 & 2.50 & 1.72 \\
\cline{2-12}
& \multirow{3}{*}{*} & Eq.\eqref{eq:16} & * & * & * & - & - & - & - & - & - \\
\cline{3-12}
& & GC & - & - & - & * & * & * & * & * & * \\
\cline{3-12}
& & RB & - & - & - & * & * & * & * & * & * \\
\hline
\multirow{6}{*}{\textsuperscript{138}Ba} &\multirow{3}{*}{4-50 (RIPL-3) + res.} & Eq.\eqref{eq:16} & 0.73 & 0.77 & 1.65 & - & - & - & - & - & - \\
\cline{3-12}
& & GC & - & - & - & 10.51 & 1.21 & 24.46 & 3.11 & 1.09 & 28.22 \\
\cline{3-12}
& & RB & - & - & - & 6.33 & -1.40 & 29.93 & 3.41 & 0.09 & 30.21 \\
\cline{2-12}
& \multirow{3}{*}{8-21 (Ref.\cite{koning2008}) + res.} & Eq.\eqref{eq:16} & 0.73 & 0.70 & 0.52 & - & - & - & - & - & - \\
\cline{3-12}
& & GC & - & - & - & 11.21 & 1.23 & 0.62 & 6.33 & 2.26 & 0.67 \\
\cline{3-12}
& & RB & - & - & - & 13.22 & 1.30 & 0.72 & 8.44 & 2.20 & 1.04 \\
\hline
\multirow{6}{*}{\textsuperscript{139}La} &  \multirow{3}{*}{4-33 (RIPL-3) + res.} & Eq.\eqref{eq:16} & 0.79 & -0.76 & 7.40 & - & - & - & - & - & - \\
\cline{3-12}
& & GC & - & - & - & 11.66 & 0.15 & 15.85 & 6.86 & 1.21 & 25.07 \\
\cline{3-12}
& & RB & - & - & - & 12.94 & 0.12 & 18.95 & 8.27 & 1.10 & 33.79 \\
\cline{2-12}
& \multirow{3}{*}{4-21 (Ref.\cite{koning2008}) + res.} & Eq.\eqref{eq:16} & 0.80 & -0.89 & 0.94 & - & - & - & - & - & - \\
\cline{3-12}
& & GC & - & - & - & 11.64 & 0.01 & 0.92 & 7.08 & 1.08 & 1.24 \\
\cline{3-12}
& & RB & - & - & - & 12.94 & -0.03 & 0.98 & 8.42 & 0.95 & 1.52 \\
\hline
\multirow{6}{*}{\textsuperscript{166}Ho} &  \multirow{3}{*}{11-154 (RIPL-3) + res.} & Eq.\eqref{eq:16} & 0.61 & -2.07 & 42.44 & - & - & - & - & - & - \\
\cline{3-12}
& & GC & - & - & - & 9.53 & -2.66 & 84.47 & 5.21 & -1.39 & 84.83 \\
\cline{3-12}
& & RB & - & - & - & 9.76 & -3.13 & 86.47 & 5.73 & -1.89 & 86.53 \\
\cline{2-12}
& \multirow{3}{*}{2-16 (Ref.\cite{koning2008}) + res.} & Eq.\eqref{eq:16} & 0.56 & -1.49 & 0.79 & - & - & - & - & - & - \\
\cline{3-12}
& & GC & - & - & - & 17.76 & -0.79 & 0.49 & 10.96 & -0.08 & 0.43 \\
\cline{3-12}
& & RB & - & - & - & 19.42 & -0.80 & 0.46 & 12.09 & -0.14 & 0.49 \\
\hline
\multirow{6}{*}{\textsuperscript{198}Au} & \multirow{3}{*}{13-91 (RIPL-3) + res.} & Eq.\eqref{eq:16} & 0.60 & -1.65 & 22.34 & - & - & - & - & - & - \\
\cline{3-12}
& & GC & - & - & - & 15.94 & -0.79 & 60.16 & 9.70 & 0.085 & 99.79 \\
\cline{3-12}
& & RB & - & - & - & 18.47 & -0.70 & 84.45 & 12.38 & 0.11 & 156.17 \\
\cline{2-12}
& \multirow{3}{*}{3-17 (Ref.\cite{koning2008}) + res.} & Eq.\eqref{eq:16} & 0.60 & -1.62 & 4.37 & - & - & - & - & - & - \\
\cline{3-12}
& & GC & - & - & - & 15.68 & -0.96 & 3.69 & 9.50 & -0.15 & 3.11 \\
\cline{3-12}
& & RB & - & - & - & 18.08 & -0.93 & 3.43 & 12.06 & -0.19 & 2.67 \\
\hline
\multirow{6}{*}{\textsuperscript{233}Th} &  \multirow{3}{*}{3-65 (RIPL-3) + res.} & Eq.\eqref{eq:16} & 0.42 & -1.05 & 35.74 & - & - & - & - & - & - \\
\cline{3-12}
& & GC & - & - & - & 25.15 & -0.40 & 131.05 & 13.50 & 0.006 & 507.50 \\
\cline{3-12}
& & RB & - & - & - & 27.88 & -0.38 & 159.46 & 16.54 & 0.006 & 614.39 \\
\cline{2-12}
& \multirow{3}{*}{7-16 (Ref.\cite{koning2008}) + res.} & Eq.\eqref{eq:16} & 0.42 & -1.07 & 0.26 & - & - & - & - & - & - \\
\cline{3-12}
& & GC & - & - & - & 24.59 & -0.55 & 0.38 & 15.13 & -0.04 & 0.52 \\
\cline{3-12}
& & RB & - & - & - & 27.24 & -0.54 & 0.41 & 18.26 & -0.07 & 0.63 \\
\hline
\end{tabular}
}
\caption{Fitting parameters of LD models obtained by $\chi^2$ minimizing (Eq.\eqref{eq:22}) of discrete and resonance data for the two sets of data from Refs. \cite{ripl3} and \cite{koning2008}. (*  data for the \textsuperscript{116}Sn in the work \cite{koning2008} does not exist.)}
\label{table:3}
\end{table}
\begin{figure}[!htbp] %ht here top demek
\centering
\subfigure[]{\includegraphics[width=0.40\linewidth]{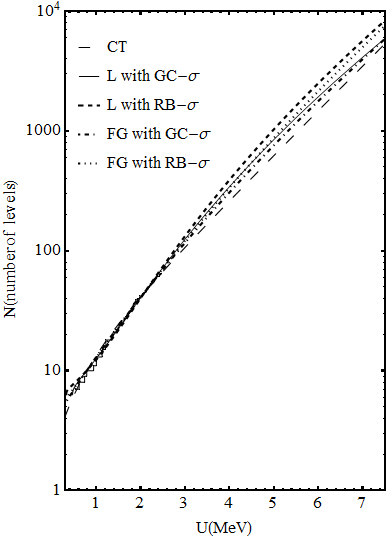}}
\label{fig:3a}
\quad
\subfigure[]{\includegraphics[width=0.40\linewidth]{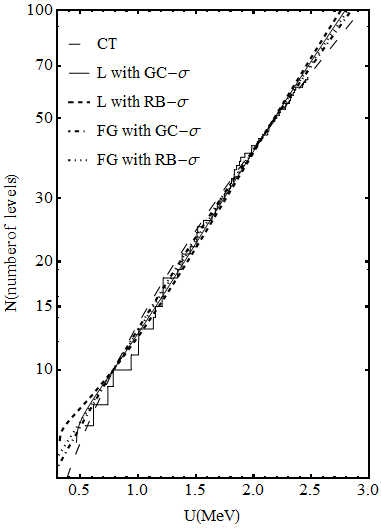}}
\label{fig:3b}
\caption{Calculated number of levels of \textsuperscript{60}Co. Model parameters are taken from Table 4. Histogram data are taken from RIPL-3. a) discrete and resonance energy range, b) discrete energy range.}
\label{fig:3}
\end{figure}
\begin{figure}[!htbp] %ht here top demek
\centering
\subfigure[]{\includegraphics[width=0.41\linewidth]{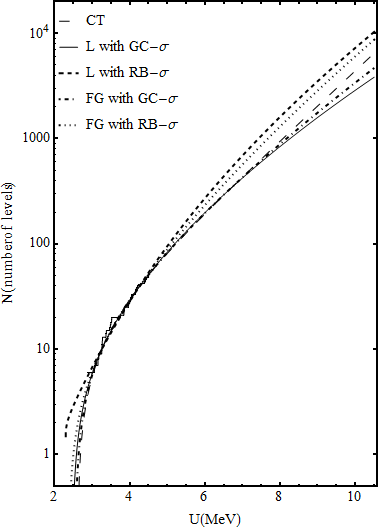}}
\label{fig:4a}
\quad
\subfigure[]{\includegraphics[width=0.40\linewidth]{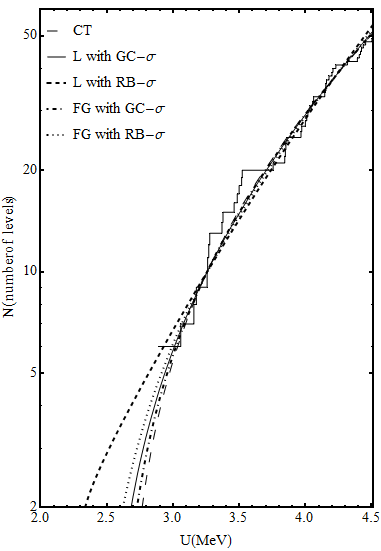}}
\label{fig:4b}
\caption{The same as in Fig. \ref{fig:3}, but for \textsuperscript{62}Ni.}
\label{fig:4}
\end{figure}

\begin{figure}[!htbp] %ht here top demek
\centering
\subfigure[]{\includegraphics[width=0.41\linewidth]{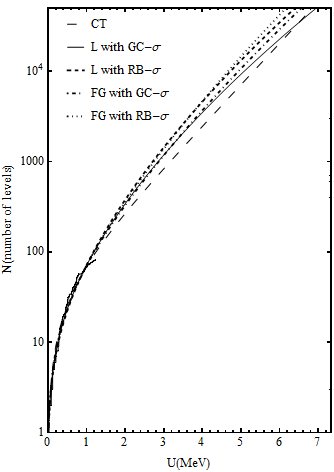}}
\label{fig:5a}
\quad
\subfigure[]{\includegraphics[width=0.40\linewidth]{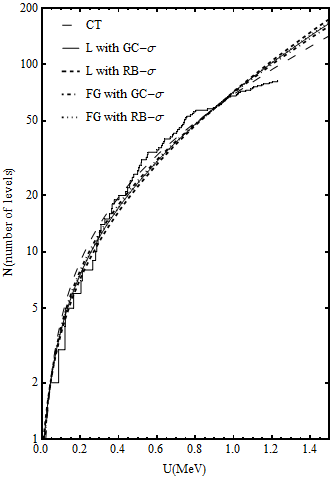}}
\label{fig:5b}
\caption{The same as in Fig. \ref{fig:3}, but for \textsuperscript{76}As.}
\label{fig:5}
\end{figure}

\begin{figure}[!htbp] %ht here top demek
\centering
\subfigure[]{\includegraphics[width=0.41\linewidth]{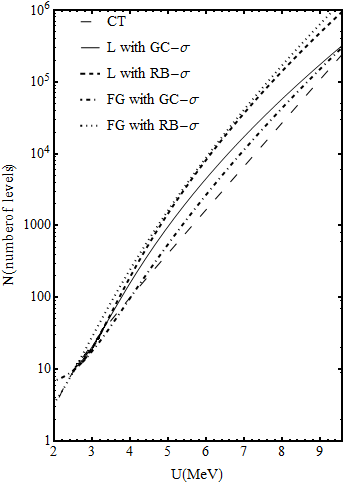}}
\label{fig:6a}
\quad
\subfigure[]{\includegraphics[width=0.40\linewidth]{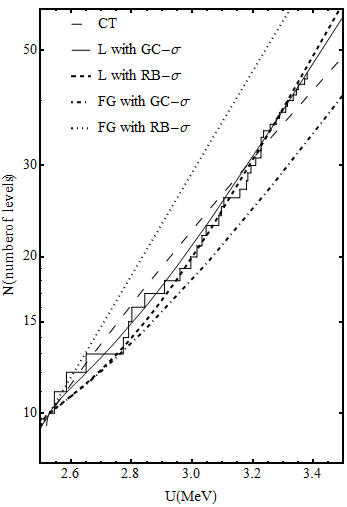}}
\label{fig:6b}
\caption{The same as in Fig. \ref{fig:3}, but for \textsuperscript{116}Sn.}
\label{fig:6}
\end{figure}
The finding values of the parameters through the models and corresponding $\chi^2$ values, finding in the global search of the parameters for the domain of the data are illustrated in Table 3. The results finding may be classified as followings: (a)The $a$ and $a_L$ parameter values for GC spin cutoff of both, L and FG, are systematically less than those for RB spin cutoff (exception is the \textsuperscript{138}Ba for the RIPL-3 data range), (b) for the range of discrete energies from Ref.\cite{koning2008} the parameter values for CT and FG are close to those of finding in \cite{egidy2005}, (c) for the both discrete energy data range in Refs.\cite{ripl3,koning2008} the LD parameter values of L and FG within the GC and RB spin cutoff's are close to each other, except for \textsuperscript{138}Ba and \textsuperscript{166}Ho, (d)for the nuclei \textsuperscript{62}Ni, \textsuperscript{76}As, \textsuperscript{166}Ho, \textsuperscript{198}Au in the range of discrete energies of work \cite{koning2008}  and for  \textsuperscript{116}Sn L theory better describes (smaller values of $\chi^2$) observed data than that do CT and FG models. The last result indicates reliability of L in describing LD's even in the range from discrete to unbound states. Figs. 3-6 show a comparison of the calculated level numbers with data for the \textsuperscript{60}Co, \textsuperscript{62}Ni, \textsuperscript{76}As and \textsuperscript{116}Sn nuclei.

\section{Concluding remarks}
\label{sec:5}

In this novel work, we have checked the applicability of the usable present form of the Landau's level density formula for describing nuclear LD's in the region of both, discrete and, discrete and s-wave neutron resonance energy range separately, and compare them with the FG and CT models. To do this, we have fitted the LD formulas of L, FG and CT with free level density and energy shift parameters for L and FG and, constant $T$ and energy shift parameters for CT to the experimental data for the nuclei  \textsuperscript{60}Co, \textsuperscript{62}Ni, \textsuperscript{76}As, \textsuperscript{116}Sn, \textsuperscript{138}Ba, \textsuperscript{139}La, \textsuperscript{166}Ho, \textsuperscript{198}Au and \textsuperscript{233}Th. Our analysis has been carried out for the range of discrete energies for the two sets of discrete energy level data from Refs. \cite{ripl3,koning2008}, used in the fitting procedure. The results obtained by $\chi^{2}$ minimizing method used in the fitting procedure has shown that the L theory reproduces better results than those of the other models when compared to data within the range of discrete energies for both data set considered. Apparently, this observation is due to the fundamental reason related to the physical conditions used the in derivation of LD expressions for L and FG. More specifically, the framework of Landau's theory for thermodynamics of system for interacting particles involves low excitations such as the energies less than the concerning neutron binding energies. In general the same observation appears when the discrete and resonance data are taken into account simultaneously with the systematic in \cite{koning2008}. The present test results for the RIPL-3 data range of discrete energies, however, indicate some advantages of the CT model in describing of LD's including s-wave neutron energy when compared to the others. The two spin cutoff parameters concerning with  GC and RB spin cutoffs, have been used in the fitting procedures for FG and L. We have observed that the fits with GC spin cutoff parameters of L have better accuracy than those of RB. 

We have also calculated the $a_{res}$ parameter values of L and FG and, the $T_{res}$ parameter values of CT model for the $66$ nuclei by the fit of corresponding LD expressions to the observed s-wave neutron resonance LD's. The results observe resonance behaviors near the doubly magic nuclei. However, we note  that the values of level density parameters ($a_L$) of L calculated for all nuclei are systematically less than those of FG. Clearly, such small values of $a_L$ of L theory lead to the decrease of LD's at higher excitations when compared to FG and CT calculations.

The present Landau model of nuclear LD's can not explicitly include different effects of nuclear structure such as shell, pairing and collective effects like FG and CT. Nevertheless, gaining confidence form the successful applications of the present model in this article, it is worth to extend the present scenario to other nuclei. 

Another way for testing the reliability of the present nuclear LD expression is the description of residual nuclei LD's extracted from the particle evaporation spectra in different nuclear reactions. Along this line the works are in progress.


\newpage

\begin{thebibliography}{1}

\bibitem{ripl3}R. Capote, M. Herman, P. Oblozinsky, P. G. Young, S. Goriley, T. Belgya, A.V. Ignatyuk, A. J. Koning, S. Hilaire, V. A. Plujko, M. Avrigeanu, O. Bersillon, M. B. Chadwick, T. Fukahori, Zhigang Ge, Yinlu Han, S. Kailas, J. Kopecky, V. M. Maslov, G. Reffo, M. Sin., E. Sh. Soukhovitskii and P. Talou, Reference Input Parameter Library (RIPL-3), Nuclear Data Sheets, 110, 12, (2009) 3107-3214 Available online at https://www-nds.iaea.org/RIPL-3/

\bibitem{koning2008}A. J. Koning, S. Hilaire, S. Goriley, Nucl. Phys. A 810 (2008) 13

\bibitem{bethe1936}H. A. Bethe, Phys. Rev. 50 (1936) 332

\bibitem{gc1965}A. Gilbert, A. G. W. Cameron, Can. J. Phys. 43 (1965) 1493

\bibitem{ignatyuk1979}A. V. Ignatyuk, K. K. Istekov, G. N. Smirenkin, Sov. J. Phys. A 217 (1979) 450

\bibitem{dilg1973}W. Dilg, w. Schantl, H. Vonach, M. Uhl, Nucl. Phys. A 217 (1973) 269

\bibitem{thomson1963}D. B. Thomson, Phys. Rev. 129 (1963) 1649 

\bibitem{malyshev1969}A. V. Malyshev, Sov. Phys. JETP 18 (1969) 221; \textit{Nuclear Structure and Level Density}, Atomizdat, Moscow, 1969

\bibitem{maruyama1969}M. Maruyama, Nucl. Phys. 131 (1969) 145

\bibitem{grimes1972}S. M. Grimes, J. D. Anderson, J. W. McClurie, B. A. Pohl, and C. Wong Phys. Rew. C, 6, 1, (1972) 236-248

\bibitem{voinov2007}A. V. Voinov, S. M. Grimes, C. R. Brune, M. J. Hornish, T. N. Massey, A. Salas, Phys. Rev. C 76 (2007) 044602

\bibitem{voinov2009}A. V. Voinov, B. M. Oginni, S. M. Grimes, C. R. Brune, M. Guttormsen, A. C. Larsen, T. N. Massey, A. Schiller, S. Siem, Phys. Rev. C 79 (2009) 031301

\bibitem{voinov2013}A. V. Voinov, S. M. Grimes, C. R. Brune, A. B\"{u}rger, A. G\"{o}rgen, M. Guttormsen, A. C. Larsen, T. N. Massey, S. Siem, Phys. Rev. C 88 (2013) 054607

\bibitem{ignatyuk1993}A. V. Ignatyuk, J. L. Weil, S. Raman, S. Kahane, Phys. Rev. C 47 (1993) 1504

\bibitem{hilaire2006}S. Hilaire, S. Goriley, Nucl. Phys. A 779 (2006) 63

\bibitem{landau1937}L. D. Landau, Sov. J. JETP 7 (1937) 819

\bibitem{egidy2005}T. von Egidy, D. Bucurescu, Phys. Rev. C 72 (2005) 044311; T. von Egidy and D. Bucurescu, Phys. Rev. C 80 (2009) 054310 

\bibitem{baba1970}H. Baba, Nucl. Phys. A 159 (1970) 625



 
\end{thebibliography}
\end{document}